\documentclass[pdftex,twocolumn,epjc3]{svjour3}
\pdfoutput=1

\usepackage[T1]{fontenc}

\smartqed  % flush right qed marks, e.g. at end of proof

\RequirePackage[colorlinks,citecolor=blue,urlcolor=blue,linkcolor=blue]{hyperref}

\usepackage[all, warning]{onlyamsmath}
\usepackage{amssymb, amsmath}
\usepackage{mathptmx}
\usepackage{graphicx}
\usepackage{dcolumn}
\usepackage{bm}
\usepackage[figuresright]{rotating}
\usepackage{fancyhdr}
\usepackage{enumitem}
\usepackage{siunitx}
\usepackage{indentfirst}
\usepackage{xspace}
\usepackage{grffile}
\usepackage[utf8]{inputenc}
\usepackage{color}
\usepackage{tabularx}
\usepackage{microtype}

\newcommand{\pp}{p+p}
\newcommand{\dAu}{d+\text{Au}}
\newcommand{\pAu}{p+\text{Au}}
\newcommand{\pPb}{p+\text{Pb}}
\newcommand{\pA}{p+A}
\newcommand{\gamp}{\gamma+p}
\newcommand{\gamd}{\gamma+d}
\newcommand{\gamA}{\gamma+A}
\newcommand{\pizero}{\pi^{0}}
\newcommand{\jpsi}{J/\psi}
\newcommand{\pt}{p_\text{T}}
\newcommand{\ptmin}{p_\text{T\,min}}
\newcommand{\pz}{p_\text{z}}
\newcommand{\wgp}{W_{\gamma+p}}
\newcommand{\snn}{\sqrt{s_{NN}}}
\newcommand{\ylab}{y_\text{lab}}
\newcommand{\etalab}{\eta_\text{lab}}
\newcommand{\mev}[1]{\SI{#1}{\mega\electronvolt}}
\newcommand{\gev}[1]{\SI{#1}{\giga\electronvolt}}
\newcommand{\tev}[1]{\SI{#1}{\tera\electronvolt}}
\newcommand{\egam}{E_\gamma}
\newcommand{\ecut}{E_\gamma^\textrm{cut}}
\newcommand{\etal}{\textrm{et al}.}

\journalname{Eur. Phys. J. C}

\begin{document}\sloppy

\title{Forward hadron production in ultra-peripheral proton--heavy-ion collisions
at the LHC and RHIC}

\author{Gaku Mitsuka \thanksref{e1,addr1}}

\thankstext{e1}{e-mail: gaku.mitsuka@cern.ch}
\institute{Universit\`a degli Studi di Firenze and INFN Sezione di Firenze,\\
           Via Sansone 1, 50019 Sesto Fiorentino (Fi), Italy \label{addr1}}

%\date{Received: date / Accepted: date}
\date{\today}

%\linenumbers

\maketitle

%
%---- Abstract ----
%
\begin{abstract}
We present a hadron production study in the forward rapidity region in
ultra-peripheral proton--lead ($\pPb$) collisions at the LHC and proton--gold
($\pAu$) collisions at RHIC.
The present paper is based on the Monte Carlo simulations of the interactions of
a virtual photon emitted by a fast moving nucleus with a proton beam.
The simulation consists of two stages: the \textsc{starlight} event generator
simulates the virtual photon flux, which is then coupled to the \textsc{sophia},
\textsc{dpmjet}, and \textsc{pythia} event generators for the simulation of
particle production.
According to these Monte Carlo simulations, we find large cross sections for
ultra-peripheral collisions particle production, especially in the very forward
region.
We show the rapidity distributions for charged and neutral particles, and the
momentum distributions for neutral pions and neutrons at high rapidities.
These processes lead to substantial background contributions to the
investigations of collective nuclear effects and spin physics.
Finally we propose a general method to distinguish between proton--nucleus
($\pA$) inelastic interactions and ultra-peripheral collisions which implements
selection cuts based on charged-particles multiplicity at mid-rapidity and/or
neutron activity at negative forward rapidity.
\end{abstract}

%
%---- Introduction ----
%
\section{Introduction}
\label{sec:introduction}

High-energy $\pA$ collisions can be classified into the following two categories
depending on the impact parameter $b$.
In the first category, $\pA$ collisions occur with geometrical overlap of the
colliding proton and nucleus, where the impact parameter is smaller than the sum
of the radii of each particle, namely, $b < R_p + R_A$ ($R_p$ and $R_A$ are the
radius of the proton and nucleus, respectively.)

In the second category instead, the impact parameter exceeds the sum of the two
radii, $b > R_p + R_A$, thus there is no geometrical overlap between the
colliding hadrons and hadronic interactions are strongly suppressed.
Nevertheless, virtual photons emitted from one of the two colliding hadrons may
anyway interact with another hadron. This process is usually referred to as
ultra-peripheral collision (UPC, see Ref.~\cite{Bertulani1,Bertulani2} for a
review).

UPCs, so far, have been used for the determination of the gluon distribution in
protons and nuclei.
For example, photoproduction of quarkonium in ultra-peripheral $\pA$ collisions
can probe a high, or possibly saturated, parton density in protons at small
Bjorken-$x$ (i.e., small parton momentum fraction of the momentum of protons).
Indeed measurements already exist of exclusive $\jpsi$ photoproduction at the
CERN Large Hadron Collider (LHC), namely, $p + \textrm{Pb} \to p + \textrm{Pb} +
\jpsi$~\cite{ALICE}.
Conversely, less attention has been paid, in UPCs, to particle production in
general photon--proton interactions, i.e., $\gamp \to X$, but nevertheless such
particle production should be considered as well in the investigation of
collective nuclear effects. Because a large cross section is expected, this
process in UPCs provides significant background events to pure $\pA$ inelastic
interaction events (hereafter ``hadronic interaction'', unless otherwise noted)
used for such investigations.
Indeed, a sizable cross section was found for hadron production in
ultra-peripheral $\dAu$ collisions~\cite{Guzey}, which amounted to
$\sim\SI{10}{\percent}$ of the $\dAu$ inelastic cross section. However, in
Ref.~\cite{Guzey}, only the cross section for UPCs was presented, and the
discussion of the rapidity and momentum distributions of the UPC induced events
was unfortunately neglected.

In this paper, we discuss the effects of particle production by $\gamp$
interaction in ultra-peripheral $\pA$ collisions compared to the measurements of
hadronic interactions in terms of the rapidity and momentum distributions,
especially in forward rapidity regions at the LHC and the BNL Relativistic Heavy
Ion Collider (RHIC).
Concerning $\pPb$ collisions at $\snn = \tev{5.02}$ at the LHC, we perform the
calculations assuming that the measurements of $\pizero$s and neutrons are made
with zero-degree calorimeters (ZDCs, for example, the
ATLAS-ZDCs~\cite{ATLASTDR}) and the LHCf detector~\cite{LHCfTDR}, which are
capable of investigating nuclear effects using hadronic interaction events in
the very forward region.
For the case at RHIC, we consider the $\pizero$ and neutron measurements in
$\pAu$ collisions at $\snn = \gev{200}$ in the year 2015. The STAR and PHENIX
experiments propose a study on partonic processes in nuclei using forward prompt
photons, where decay photons from $\pizero$s (from both the hadronic interaction
and UPCs) would be the dominant background events~\cite{STARNOTE,PHENIXMPC}.
Furthermore, measurements of the hadronic-interaction-induced prompt photons and
$\pizero$s in transversely polarized $\pAu$ collisions may provide key
information on the yet unestablished contributions of Sivers and Collins effects
to the single spin asymmetry~\cite{STARNOTE,PHENIXMPC}.

Our quantitative discussions on forward hadron production are based on Monte
Carlo (MC) simulations.
The MC simulation for UPCs consists of two steps; the virtual photon flux is
simulated by the \textsc{starlight} event
generator~\cite{STARLIGHT,STARLIGHTcode} and then the subsequent particle
production in $\gamp$ interactions is simulated by the
\textsc{sophia}~\cite{SOPHIA,SOPHIAcode},
\textsc{dpmjet}~\cite{DPMJET,DPMJETcode}, and \textsc{pythia}~\cite{PYTHIA,PYTHIAcode}
event generators. \textsc{starlight} in this study has been partially customized
in order to transfer the information on the simulated virtual photon to
\textsc{sophia}. The MC simulation for hadronic interactions is performed by the
\textsc{dpmjet} alone.

The paper is organized as follows. First, in Sect.~\ref{sec:simulation}, the
methodology of the MC simulations is explained. Next, in
Sect.~\ref{sec:results}, we discuss the simulation results in terms of the
rapidity and momentum distributions, where the hadron production in UPCs is
compared to that in hadronic interactions. Additionally, we attempt a reduction
in UPC events by requiring associated particles. Conclusions are drawn in the
last section. In this paper natural units $\hbar = c = 1$ are used throughout.

%
%---- MC simulation ----
%
\section{Monte Carlo simulations Methodology}
\label{sec:simulation}

As stated above, the MC simulation for UPCs in this study consists of two steps.
First, we simulate the virtual photon flux as a function of the photon energy
and impact parameter by using \textsc{starlight}~\cite{STARLIGHT}. Next, the
simulation of the $\gamp$ interaction is performed by using
\textsc{sophia}~\cite{SOPHIA} at low energy and either
\textsc{dpmjet}~\cite{DPMJET} or \textsc{pythia}~\cite{PYTHIA} at high energy.
The methodology of the MC simulation for UPCs is explained in the following
subsections from \ref{sec:photon} to \ref{sec:highe}.
In these subsections, the proton rest frame is referenced to unless otherwise
noted.
The MC simulation for hadronic interactions is simply performed by using
\textsc{dpmjet}, and will be described in Sect.~\ref{sec:inelastic}.

\subsection{Virtual photon flux simulation}
\label{sec:photon}

In this paper, the energy spectrum of the virtual photons emitted by the
relativistic nucleus follows the Weizs\"{a}cker-Williams
approximation~\cite{Weizsacker,Williams} implemented in \textsc{starlight}.
The double differential photon flux due to the fast moving nucleus with velocity
$\beta$ is written as
\begin{equation}
\frac{d^3N}{d\egam db^2} = \frac{Z^2\alpha}{\pi^2}\frac{x^2}{\egam b^2}
\left( K_1^2(x) + \frac{1}{\gamma^2}K_0^2(x) \right),
\label{eq:doublediff}
\end{equation}
where $N$ is the number of the emitted photons, $\egam$ is the photon energy,
$Z$ is the electric charge ($Z = 82$ for Pb and $Z = 79$ for Au), $\alpha$ is
the fine structure constant, $x = \egam b/\gamma$ ($\gamma =
\sqrt{1-\beta^2}^{-1/2}$ is the Lorentz factor), and $K_0$ and $K_1$ are the
modified Bessel functions. In the case of a relativistic nucleus ($\gamma \gg
1$), the contribution of the term $K_0^2(x)/\gamma^2$ in
Eq.~(\ref{eq:doublediff}) can be safely disregarded, and in fact
\textsc{starlight} considers only the term $K_1^2(x)$.
For heavy nuclei with a large radius, the virtuality of the photon
$|q^2|<(1/R_A)^2$ can be neglected. Thus all photons are treated as real photons
in the simulation for this analysis.
Another approximation is due to the fact that here we assume a point charge for
the nucleus and this assumptions may lead to a certain level of systematic
uncertainty.
For example, as discussed in Ref.~\cite{Gawenda}, the photon flux in reality
depends on the choice of the form factor in the nucleus by $\lesssim
\SI{20}{\percent}$.

The probability $P_{\textrm{UPC}(\gamp \to X)}(b)$ for a single photon
interaction with a proton in UPCs as a function of $b$ is given by
\begin{equation}
P_{\textrm{UPC}(\gamp \to X)}(b) =
\int_{\egam^\textrm{min}}^{\egam^\textrm{max}} \frac{d^3N}{d\egam db^2}
\sigma_{\gamp \to X}(\egam) \overline{P_\textrm{had}}(b)\,d\egam
\label{eq:pupc}
\end{equation}
where $\sigma_{\gamp \to X}(\egam)$ is the total cross section for a single real
photon interaction with a rest proton and $\overline{P_\textrm{had}}(b)$ is the
probability of having no hadronic interactions in $\pA$ collisions.
$\egam^\textrm{min}$ and $\egam^\textrm{max}$ are the minimum and maximum photon
energies.

In this study, we take $\sigma_{\gamp \to X}(\egam)$ from the compilation of
present experimental results~\cite{PDG} when a photon--proton center-of-mass
energy $\wgp$ is smaller than $\gev{7}$.
A linear interpolation is performed between each data point. The cross section
at the exact photopion production threshold, $\egam=\gev{0.15}$, for which no
experimental measurement exists, is forced to zero.
At $\wgp$ larger than $\gev{7}$, $\sigma_{\gamp \to X}(\egam)$ is derived from
the best COMPETE fit results~\cite{PDG}.

A finite probability for having no hadronic interactions
$\overline{P_\textrm{had}}(b)$ is introduced in order to implement a smooth cut
off for values of the impact parameter approaching: $b=R_p + R_A$.
$\overline{P_\textrm{had}}(b)$ is calculated from the Woods-Saxon nuclear
density and the Glauber model~\cite{STARLIGHT}.

The range of the impact parameter $b$ considered in the simulation extends from
$b^\textrm{min} = \SI{4}{\femto\meter}$ to $b^\textrm{max} =
\SI{e5}{\femto\meter}$. $b^\textrm{min}$ is well below the sum of the effective
radii of colliding particles ($\sim\SI{8}{\femto\meter}$ for both $\pPb$ and
$\pAu$ collisions), and $\overline{P_\textrm{had}}(b)$ rapidly approaches zero
below $\SI{8}{\femto\meter}$.
The photon energy $\egam$ in the simulation ranges from slightly above the
photopion production threshold, i.e., $\egam^\textrm{min} = \gev{0.16}$, to
$\egam^\textrm{max}$.
$\egam^\textrm{max}$ is obtained from $\gamma/b^\textrm{min}$ and amounts to
$\tev{700}$ for $\pPb$ collisions at LHC and $\tev{1.1}$ for $\pAu$
collisions at RHIC.

\subsection{Simulation of the low-energy photon--proton interaction}
\label{sec:lowe}

The particle production from the interaction of a low-energy photon with a
proton is simulated by the \textsc{sophia} 2.1 event generator~\cite{SOPHIA}.
In \textsc{sophia}, particle production via baryon resonances, direct pion
production, and multiparticle production are taken into account. For the baryon
resonances, the known resonances from $\Delta(1232)$ to $\Delta(1950)$ are
considered with their physical parameters. The resonance decays isotropically,
depending on the available phase space.
The non-diffractive interaction, which is implemented using the dual parton
model~\cite{DPM}, starts to dominate at $\egam \gtrsim \gev{2}$ increasing with
energy.
The diffractive interaction is implemented as the quasi-elastic exchange of a
reggeon or pomeron between virtual hadronic states of the photon and the proton.
The \textsc{sophia} generator is used for the UPC simulations, with the photon
energy ranging from $\egam^\textrm{min}$ to $\ecut$.
$\ecut$ is a "technical" cut off energy that distinguishes low energy from
high-energy interactions for \textsc{sophia} and the other generators.

Here we emphasize that, firstly, the simulation with the photon energy $\egam
\lesssim \gev{0.5}$ is crucial for producing low transverse-momentum ($\pt$) UPC
induced events that are dominant in the very forward regions of the detector
reference frame (explained in Sect.~\ref{sec:momentum}), secondly,
\textsc{sophia} can simulate the interaction of such a low-energy photon with a
proton above the photopion production threshold, and, finally, a newly developed
interface to \textsc{sophia} has been introduced into \textsc{starlight} that
was not originally coupled to \textsc{sophia}. This interface provides two main
functions: first, the information on the simulated photon by \textsc{starlight}
is transferred to \textsc{sophia}, and second, the information on the produced
particles after the simulation of the $\gamp$ interaction are returned from
\textsc{sophia} to \textsc{starlight} for, e.g., Lorentz boost, listing of the
produced particles, etc.

\subsection{Simulation of high-energy photon--proton interaction}
\label{sec:highe}

At the photon energy $\egam > \ecut$, we perform the simulation of $\gamp$
interactions by using either \textsc{pythia} 6.428 or \textsc{dpmjet} 3.05.
\textsc{starlight} has its own interface to both event generators.

In \textsc{pythia}~\cite{PYTHIA}, the high-energy photon interactions with a
proton are classified into three different schemes~\cite{Schuler}. Direct events
describe the bare photon interaction with a parton from the proton, typically
leading to high $\pt$ jets.
In vector meson dominance (VMD) events, the photon fluctuates into a vector
meson and then the vector meson interacts with the proton. This class includes
low-$\pt$ events.
Finally, generalized VMD events are where the photon fluctuates into a
$q\bar{q}$ pair which interacts with a parton from the proton.
Single photon dissociation and single proton dissociation occur in the
relatively low $\pt$ region.
\textsc{pythia} requires a simulated event that has a center-of-mass energy
$\wgp$ larger than $\gev{10}$. This energy corresponds to a photon energy
$\egam=\gev{55}$. Thus \textsc{sophia} and \textsc{pythia} are employed for the
simulation of a $\gamp$ interaction for photon energies below and above
$E_\gamma^{\textrm{cut},\textsc{pythia}} = \gev{55}$, respectively.

\textsc{dpmjet}~\cite{DPMJET} is based on the two-component dual parton model.
$\gamp$ interactions are especially implemented in the \textsc{phojet} MC event
generator~\cite{PHOJET} inside \textsc{dpmjet}. In \textsc{phojet}, the physical
photon is described as a superposition of the bare photon and virtual hadronic
photon. The bare photon directly interacts with partons from the proton. The
virtual hadronic photon first fluctuates into a $q\bar{q}$ pair and then
hadronically interacts with the proton.
Both single photon dissociation and single proton dissociation are also taken
into account.
\textsc{dpmjet} requires $E_\gamma^{\textrm{cut},\textsc{dpmjet}} \geq \gev{6}$,
the lowest energy that guarantees usable results from the model.
Thus, \textsc{sophia} and \textsc{dpmjet} are employed for the simulation of a
$\gamp$ interaction for photon energies below and above $\gev{6}$, respectively.

As summarized in Table~\ref{tbl:generators}, we have thus two types of UPC
simulations: the first one given by the simulations of photohadron production
performed by \textsc{sophia} and \textsc{pythia} with a cut off energy of
$E_\gamma^{\textrm{cut},\textsc{pythia}} = \gev{55}$, the second one deriving
from simulations performed by \textsc{sophia} and \textsc{dpmjet} with
$E_\gamma^{\textrm{cut},\textsc{dpmjet}} = \gev{6}$.

\begin{table}[tbp]
    \caption{Summary of the event generators for $\gamp$ interactions and their
    cut off energies. ``Low-energy'' and ``High-energy'' in the table
    indicate the energy regions $\egam^\textrm{min} < \egam < \ecut$ and
    $\ecut < \egam < \egam^\textrm{max}$, respectively.}
\begin{tabular}{lccc}
\hline\hline
     & \multicolumn{2}{c}{$\gamp$ interactions} & \\
     \cline{2-3}
     & Low-energy & High-energy & $\ecut$ \\ 
     \hline
SOPHIA+PYTHIA & \textsc{sophia} 2.1 & \textsc{pythia} 6.428 & $\gev{55}$ \\
SOPHIA+DPMJET & \textsc{sophia} 2.1 & \textsc{dpmjet} 3.05  & $\gev{6}$  \\
\hline\hline
\end{tabular}
\label{tbl:generators}
\end{table}

\subsection{Simulation of hadronic interactions}
\label{sec:inelastic}

In this paper, \textsc{dpmjet} is used as an event generator for the MC
simulation of hadronic interactions, which include non-diffractive and
diffractive interactions but do not include elastic scattering.
The multiple scattering process in the interaction with a nuclear target, which
causes nuclear shadowing, is described by the Gribov-Glauber
model~\cite{Gribov,Glauber} in terms of the multiple pomeron exchange. Some of
the parameters for soft particle production are set at the values that best
reproduce experimental results.
The integrated interface \textsc{crmc} 1.5.3~\cite{CRMC} is used to access the
\textsc{dpmjet} generator.

%
%---- Simulation results ----
%
\section{Predictions for ultra-peripheral collisions at LHC and RHIC}
\label{sec:results}
%
%---- Cross section ----
%
\subsection{Total cross sections}
\label{sec:crosssection}

The total cross section for UPCs ($\sigma_{\text{UPC}(\gamp \to X)}$) is
calculated by integrating Eq.~(\ref{eq:pupc}) over the parameter $b$:
\begin{equation}
\begin{split}
\sigma_{\text{UPC}(\gamp \to X)} &=
\int_{b^\textrm{min}}^{b^\textrm{max}}P_{\textrm{UPC}(\gamp \to X)}\,db^2\\
&=
2\pi\int_{b^\textrm{min}}^{b^\textrm{max}}\int_{\egam^\textrm{min}}^{\egam^\textrm{max}}
\frac{d^3N}{d\egam db^2} \sigma_{\gamp \to X}(\egam)\\
&\hspace{11pt}\times \overline{P_\textrm{had}}(b)b\,db\,d\egam.
\label{eq:sigupc}
\end{split}
\end{equation}

As discussed in Sect.~\ref{sec:photon}, Eq.~(\ref{eq:doublediff}) and
$\overline{P_\textrm{had}}(b)$ are obtained by using \textsc{starlight}, while
$\sigma_{\gamp \to X}(\egam)$ is taken from experimental measurements and best
fit results to these measurements; thus, Eq.~(\ref{eq:sigupc}) is independent
from the other event generators:
\textsc{sophia}, \textsc{dpmjet}, and \textsc{pythia}. The calculated cross
sections $\sigma_{\text{UPC}(\gamp \to X)}$ at the LHC and RHIC are summarized
in Table~\ref{tbl:crosssection}. The cross sections for $\pA$ inelastic
interactions, calculated by using \textsc{dpmjet}, are also presented.
We find a sizable $\sigma_{\text{UPC}(\gamp \to X)}$s, which amount to
$\SI{20}{\percent}$ and $\SI{9}{\percent}$ of the hadronic cross sections at the
LHC and RHIC respectively. Effective cross sections that require forward
$\pizero$ and neutron tagging will be discussed later in
Sect.~\ref{sec:momentum}.

For reference we have computed the cross section in ultra-peripheral $\dAu$
collisions at $\snn=\gev{200}$ in the same manner as Eq.~(\ref{eq:sigupc}) and
using the same photon energy and impact parameter ranges as in
Ref.~\cite{Guzey}, obtaining a $\sigma_{\text{UPC}(\gamd \to X)} =
\SI{230}{\milli\barn}$, which is compatible with the result obtained by
Ref.~\cite{Guzey}.
Since the photon energy and impact parameter ranges used for the comparison are
narrower than the ranges used in the computation of
Table~\ref{tbl:crosssection}, the cross section $\sigma_{\text{UPC}(\gamd \to
X)}$ is smaller than twice the cross section $\sigma_{\text{UPC}(\gamp \to X)}$
in Table~\ref{tbl:crosssection}, namely, $2 \sigma_{\text{UPC}(\gamp \to X)} =
\SI{340}{\milli\barn}$. The factor of 2 comes from \ $\sigma_{\gamma + d \to
X}(\egam) \approx 2\sigma_{\gamp \to X}(\egam)$.

Now, to have a comparison with $\sigma_{\text{UPC}(\gamp \to X)}$, we have also
calculated the cross section $\sigma_{\text{UPC}(\gamA \to X)}$ for UPCs where a
photon emitted by a fast moving proton interacts with a nucleus $A$ ($A$ is
either Pb or Au), although this process is not taken into account in the MC
simulations performed for this study.
In ultra-peripheral $\pA$ collisions, the number of photons emitted by the
proton is generally smaller than that by the nucleus due to the $Z^2$
dependence.
The interactions of a single real photon with a nucleus can be roughly
classified into two categories in terms of the photon energy in the nucleus rest
frame, namely below or above the photopion production threshold.
For $\mev{7} < \egam < \mev{140}$, thus below the photopion production
threshold, the cross sections for photonuclear absorption processes are studied
for Pb and Au nuclei in Ref.~\cite{Vyssiere,Lepretre}. Here a certain number of
neutrons are emitted from the decay of the photoexcited nucleus in the event,
in particular at least one neutron is emitted almost $\SI{100}{\percent}$ of
the time.
For $\egam > \mev{140}$, the $\gamA$ cross section can be calculated within the
Gribov-Glauber approximation~\cite{gammaA}. The calculated $\gamA$ cross section
$\sigma_{\gamA \to X}(\egam)$ is compatible with or smaller than the simply
scaled $\gamp$ cross section $(Z+N)\sigma_{\gamp \to X}(\egam)$ ($Z$ and $N$ are
the atomic number and neutron number of the nucleus, respectively.) The
suppression of $\sigma_{\gamA \to X}(\egam)$ relative to $(Z+N)\sigma_{\gamp \to
X}(\egam)$ appears at $\egam>\gev{2}$ because of nuclear
shadowing~\cite{gammaA}.
Consequently, in ultra-peripheral $\pAu$ and $\pPb$ collisions, the process with
photon emission from the proton is suppressed with respect to that from the
nucleus.
The cross section of the former process ($\sigma_{\text{UPC}(\gamA \to X)}$)
amounts to $\SI{6}{\percent}$ of the latter ($\sigma_{\text{UPC}(\gamp \to
X)}$).

Since the UPC simulations used in this paper assume that only a single photon is
produced from the moving nucleus in any event, UPCs involving two or more
photons are not taken into account in the study. However UPC processes with the
exchange of two photons between the proton and the nucleus generally have a huge
cross section. In particular, the two-photon exchange leading to di-electrons
has a cross section $\sigma_{\text{UPC}(\gamma+\gamma \to e+e)}$ of
$\sim\SI{29}{\barn}$ at the LHC and of $\sim\SI{4}{\barn}$ at RHIC ($\propto
(Z_p\alpha)^2(Z_A\alpha)^2$, where $Z_p$ and $Z_A$ are the electric charges of
the proton and nucleus, respectively). These cross sections are obtained by
scaling the corresponding cross sections in ultra-peripheral Pb+Pb and Au+Au
collisions~\cite{Ivanov,Baltz2} with the ratio of $(Z_p/Z_A)^2$, respectively.
The two-photon exchange leading to other particle pairs, e.g., $\mu^+\mu^-$,
$\tau^+\tau^-$, and mesons, are at most $\SI{e-3}{\percent}$ compared with the
di-electron channel~\cite{Baltz3}.
Concerning forward $\pizero$ and neutron productions providing background
contribution to hadronic interaction events, we consider a higher-order process
where the two-photon exchange is accompanied by an additional inelastic
interaction of the single photon emitted from the nucleus with the proton. This
process can be factorized into two subprocesses: the two-photon exchange and the
$\gamp$ interaction~\cite{Hencken1}. Accordingly, the cross section is
calculated using:
\begin{equation}
\begin{split}
\sigma_{\text{UPC}(\gamma+\gamma \to e+e,~\gamp \to X)} &=\\
\int_{b^\textrm{min}}^{b^\textrm{max}}
P_{\textrm{UPC}(\gamma+\gamma \to e+e)}&(b)
P_{\textrm{UPC}(\gamp \to X)}(b)\,db^2,
\label{eq:twogamma}
\end{split}
\end{equation}
where $P_{\textrm{UPC}(\gamma+\gamma \to e+e)}(b)$ is the probability for the
two-photon exchange leading to di-electron and $P_{\textrm{UPC}(\gamp \to
X)}(b)$ is the probability defined in Eq.~(\ref{eq:pupc}).
With the former probability taken from the results in
Refs.~\cite{Guclu,Hencken2}, we obtain $\sigma_{\text{UPC}(\gamma+\gamma \to
e+e,~\gamp \to X)} = \SI{66}{\milli\barn}$ at the LHC and $\SI{15}{\milli\barn}$
at RHIC. These cross sections correspond to $\SI{15}{\percent}$ and
$\SI{9}{\percent}$ compared with $\sigma_{\text{UPC}(\gamp \to X)}$ shown in
Table~\ref{tbl:crosssection} at the LHC and RHIC, respectively.

\begin{table}[tbp]
	\centering
    \caption{Cross sections for particle production in ultra-peripheral
    collisions and hadronic interactions at the LHC and RHIC.}
\begin{tabularx}{208pt}{lcccccc}
\hline\hline
     &\multicolumn{3}{c}{UPC\,(mb)}   & \multicolumn{3}{c}{Hadronic
     interaction\,(mb)}  \\
     \cline{2-4}                      \cline{5-7}
     & Total    & $\pizero$ & $n$     & Inelastic & $\pizero$ & $n$    \\
\cline{1-7}
LHC  & 434      &  78       & 153     & 2189      &  91       & 125    \\
RHIC & 170      &   9       &  73     & 1851      &  67       &  35    \\
\hline\hline
\end{tabularx}
\label{tbl:crosssection}
\end{table}

%
%---- Rapidity ----
%
\subsection{Rapidity distributions}
\label{sec:rapidity}

\begin{figure}[tbp]
  \centering
  \includegraphics[width=8.5cm, keepaspectratio]{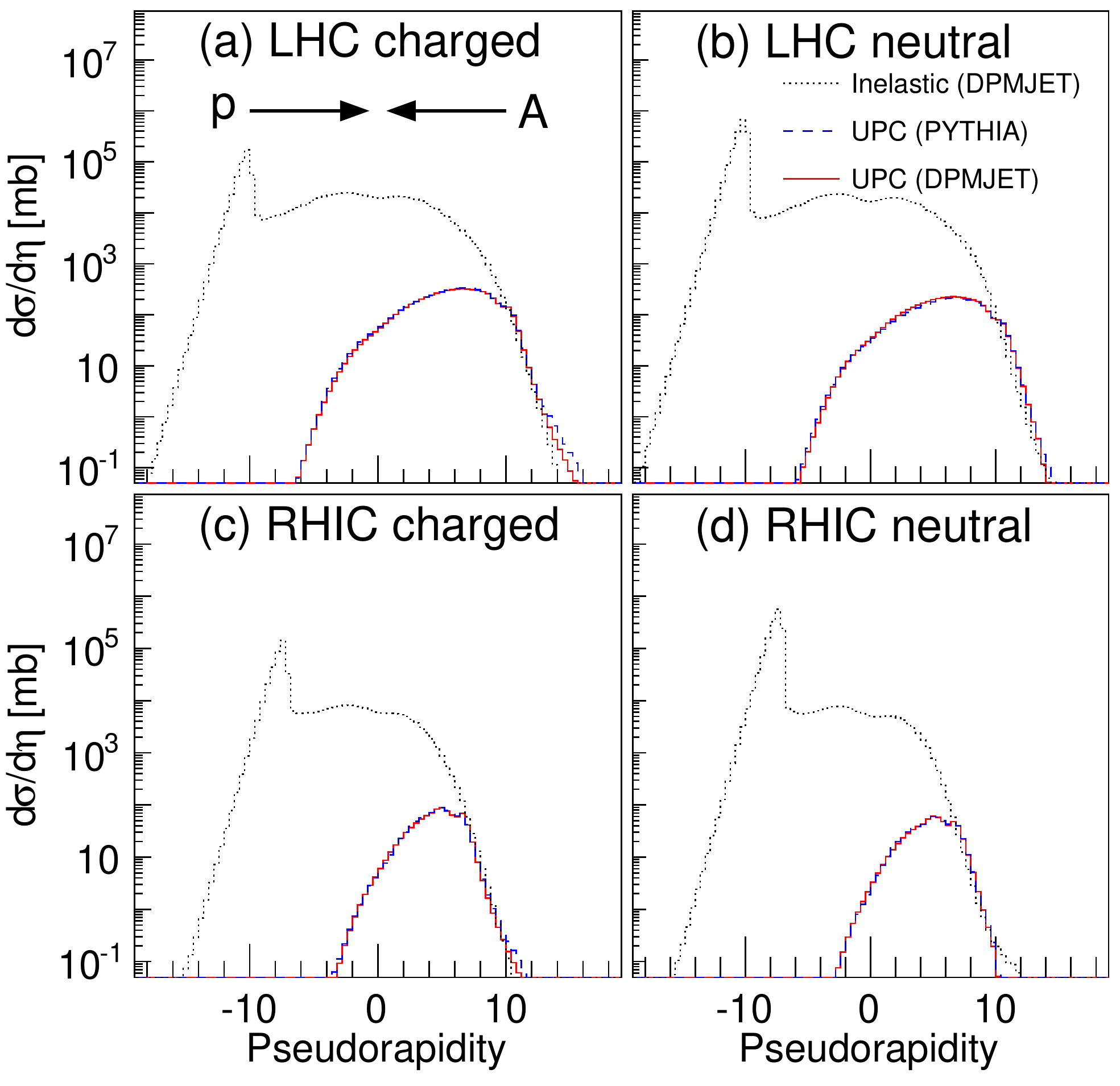}
  \caption{Charged (left) and neutral particle (right) pseudorapidity
  distributions at the LHC (top) and RHIC (bottom), respectively.
  The solid curves and dashed curves indicate the UPC simulation events
  generated by using \textsc{starlight+sophia+dpmjet} and
  \textsc{starlight+sophia+pythia}, respectively. The dotted curves indicate the
  simulated $\pPb$ and $\pAu$ inelastic events with \textsc{dpmjet} at the LHC
  and RHIC, respectively. The directions of the moving proton ($p$) and the
  nucleus ($A$) are indicated by arrows in the upper left panel.}
  \label{fig:rapidity}
\end{figure}

The charged and neutral particle pseudorapidity ($\etalab$) distributions in the
detector reference frame are shown in Fig.~\ref{fig:rapidity}. The solid curves
indicate the UPC simulation events generated by using \textsc{sophia} at low
energy and \textsc{dpmjet} at high energy. The dashed curves indicate the UPC
simulation events generated by using \textsc{sophia} at low energy and
\textsc{pythia} at high energy.
The dotted curves indicate the predictions for $\pPb$ and $\pAu$ inelastic
events with \textsc{dpmjet} at the LHC and RHIC respectively.
Hereafter the directions for the moving proton and nucleus are assumed to have
positive and negative rapidities respectively. The directions for both particles
are indicated by the arrows in the upper left panel of Fig.~\ref{fig:rapidity}
($p$ for proton and $A$ for nucleus).

In Fig.~\ref{fig:rapidity}, the pseudorapidity distributions in hadronic
interactions achieve a plateau at mid-rapidity and also have a large number of
spectator nucleons at $\etalab \sim -8$.
Conversely, particle production in UPCs is clearly clustered around positive
forward rapidities, since the UPC induced events in this study are produced only
by the interactions of the photon emitted from the nucleus with the proton.
It should be noted that the cross sections for UPCs exceed those of hadronic
inelastic interactions for charged and neutral particles at $\etalab > 9$ at the
LHC and $\etalab > 7$ at RHIC. This indicates that contamination from background
UPC events, could spoil the investigation of collective nuclear effects and spin
physics carried out from measurements of hadronic interactions in those rapidity
regions.

%
%---- Momentum distribution ----
%
\subsection{Transverse and longitudinal momentum distributions}
\label{sec:momentum}

\begin{figure*}[tbp]
  \centering
  \includegraphics[width=15cm, keepaspectratio]{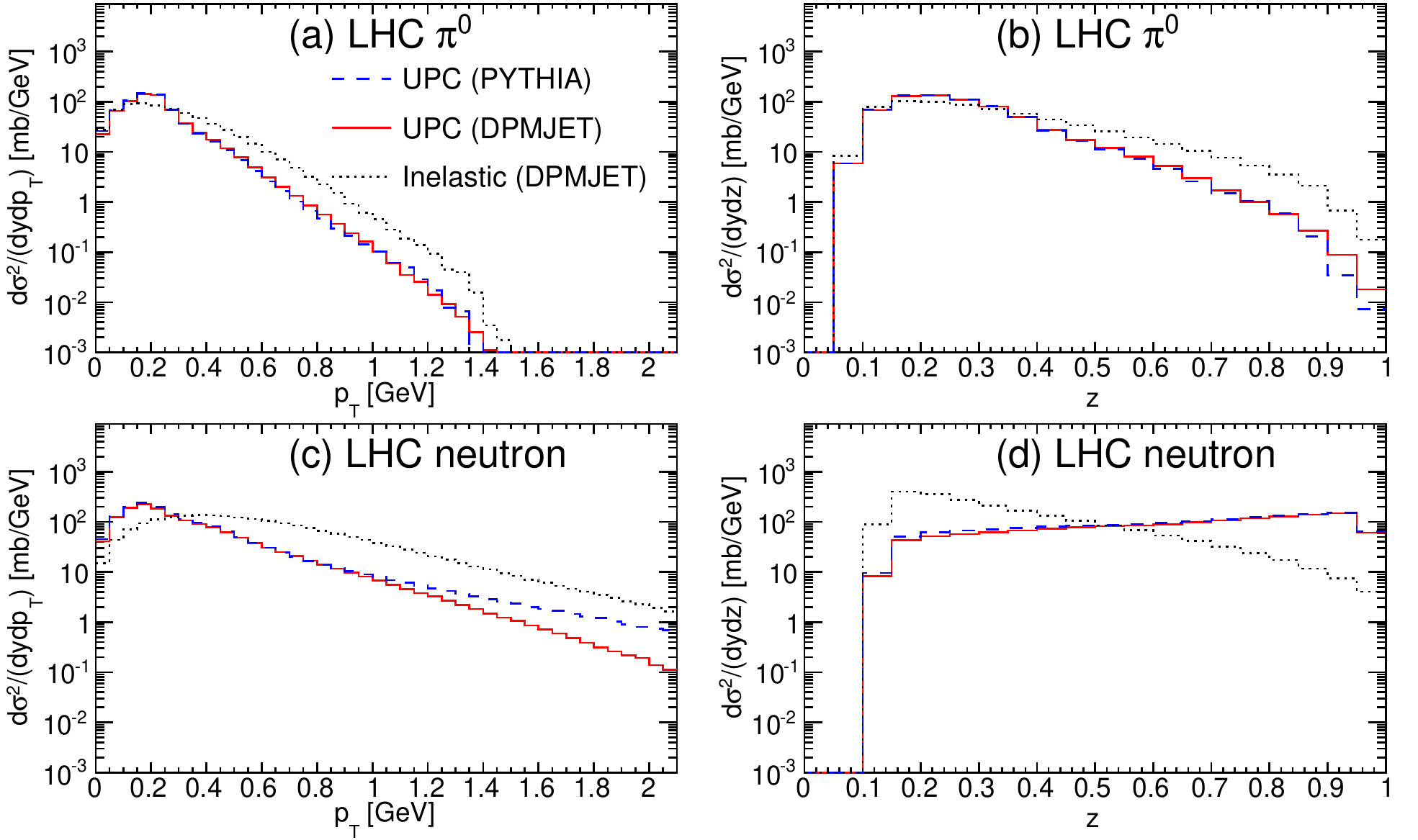}
  \caption{Simulated $\pt$ and $z$ spectra for $\pizero$s and neutrons in
  $\pPb$ collisions at the LHC. The solid curves and dashed curves indicate the
  UPC simulation events generated by using \textsc{starlight+sophia+dpmjet} and
  \textsc{starlight+sophia+pythia}, respectively. The dotted curves indicate the
  simulated $\pPb$ inelastic events with \textsc{dpmjet}.}
  \label{fig:lhc}
\end{figure*}

\begin{figure*}[tbp]
  \centering
  \includegraphics[width=15cm, keepaspectratio]{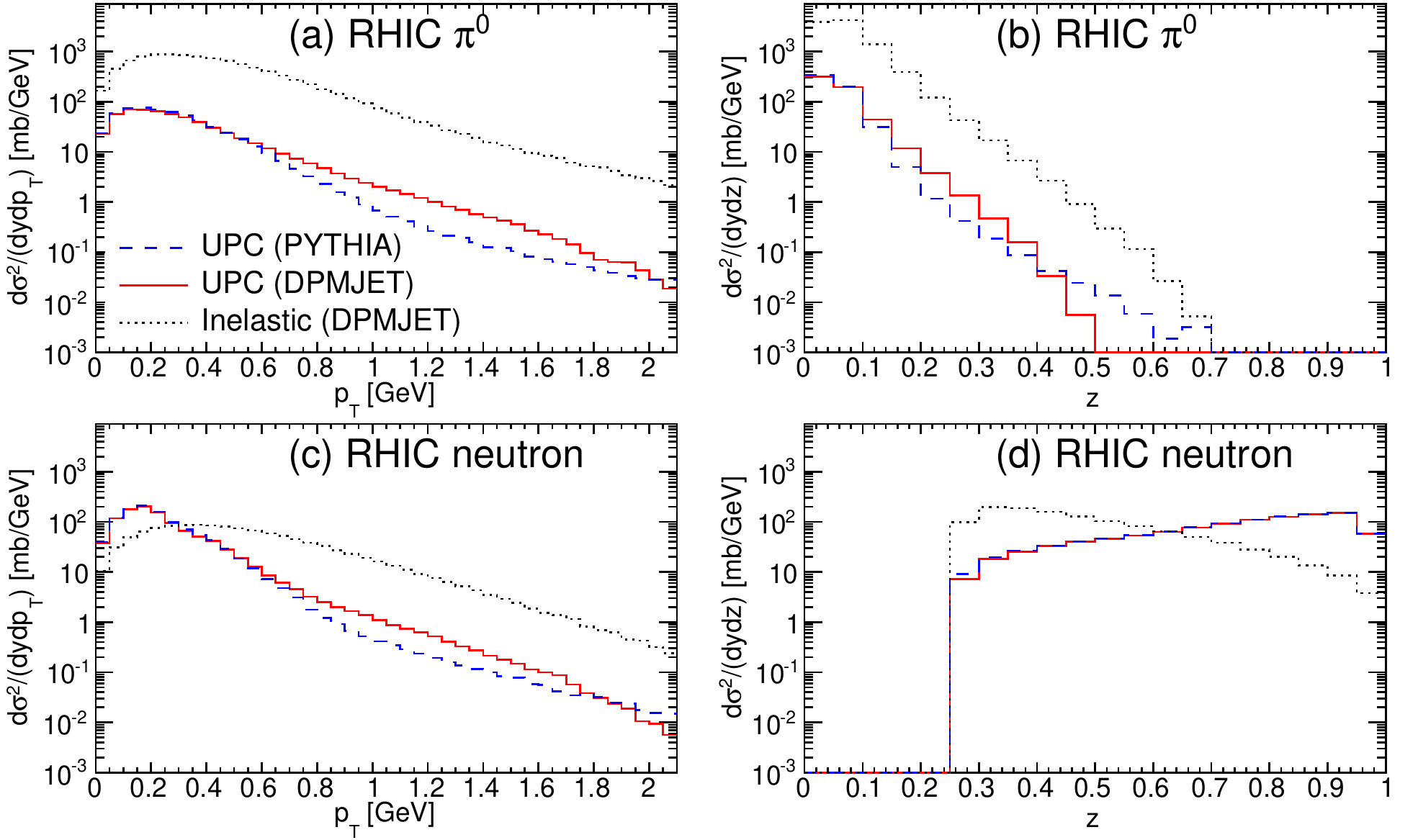}
  \caption{Simulated $\pt$ and $z$ spectra for $\pizero$s and neutrons in
  $\pAu$ collisions at RHIC. The solid curves and dashed curves indicate the UPC
  simulation events generated by using \textsc{starlight+sophia+dpmjet} and
  \textsc{starlight+sophia+pythia}, respectively. The dotted curves indicate the
  simulated $\pAu$ inelastic events with \textsc{dpmjet}.}
  \label{fig:rhic}
\end{figure*}

The simulated $\pt$ and longitudinal momentum fraction ($z$, defined as
$\pz/{p_\textrm{z~max}}$) distributions for $\pizero$s and neutrons at positive
forward rapidities (direction of the proton remnant), are shown in
Fig.~\ref{fig:lhc} for the LHC and in Fig.~\ref{fig:rhic} for RHIC.
For the distributions at the LHC, we chose rapidity regions in the detector
reference frame $8.5<\ylab<11.0$ for $\pizero$s and $7.0<\ylab<9.05$ for
neutrons, which roughly correspond to the acceptances of the
ATLAS-ZDCs~\cite{ATLASTDR} and LHCf detector~\cite{LHCfTDR}. These detectors
provide an opportunity to investigate the effects of high parton density on
forward $\pizero$s and neutrons, which emerge as a suppression in the momentum
distributions in $\pPb$ inelastic interactions relative to that of $\pp$
inelastic interactions.

In the upper left and bottom left panels of Fig.~\ref{fig:lhc}, the $\pt$
distributions of $\pizero$s and neutrons in UPCs show a steep peak at
$\pt\approx\gev{0.2}$.
These peaks originate from the $\gamp \to \pizero + p$ and $\gamp \to \pi^{+} +
n$ channels, via baryon resonances.
In fact, in the proton rest frame the $\gamp$ interactions, with a photon energy
ranging from $\egam^\textrm{min}$ to $\gev{0.5}$, have a center-of-mass energy
of $1.1 < \wgp < \gev{1.3}$ and thus occur in the baryon resonance region, which
has a larger cross section compared to other energy regions.
Conversely, the $\gamp$ interactions with higher photon energies are suppressed
due to a decrease in the photon flux with increasing photon energy.
Therefore the $\pizero$s and neutrons emitted by the decay of the baryon
resonances due to low-energy $\gamp$ interactions (dominantly $\Delta^+(1232)$)
which typically have $\pt\approx\gev{0.2}$ provide substantial contributions to
the $\pt$ distributions of each particle.
Thus, the double differential UPC cross sections exceed those of hadronic
interactions for $\pt\approx\gev{0.2}$.
The dominance of the $\gamp \to \pi^{+} + n$ channel in UPCs is also evident in
the bottom right panel of Fig.~\ref{fig:lhc}.
Forward neutrons produced in UPCs have a larger $z$ value; low momentum neutrons
produced by a low-energy $\gamp$ interaction in the proton rest frame are
boosted to nearly the same velocity of the projectile proton.
Finally, we see that the presence of UPCs certainly provides a significant
background contribution to the study of collective nuclear effects.

The $\pt$ and $z$ distributions for $\pizero$s and neutrons produced at RHIC
are shown in Fig.~\ref{fig:rhic}. We chose the rapidity regions $3.1<\ylab<3.8$
for $\pizero$s and $4.0<\ylab<5.4$ for neutrons, which correspond to
the acceptance of the MPC-EX detector~\cite{PHENIXMPC} and ZDC~\cite{RHICZDC} of
the PHENIX experiment respectively. Since these rapidity regions are near the
acceptance of the FMS+pre-shower detector~\cite{STARNOTE} and ZDC~\cite{RHICZDC}
of the STAR experiment, the following discussion is essentially
applicable also to measurements performed at the STAR experiment.
The prompt photon measurements in $\pAu$ collisions with the MPC-EX detector in
PHENIX and the FMS+pre-shower detector in STAR could provide key information on
partonic processes in the Au nucleus, whereas photons from
$\pizero$s decays, produced both from hadronic interactions and UPCs, could be the
dominant background events.
The $\pt$ distributions of $\pizero$s from UPCs have almost the same shapes as
those coming from hadronic interactions (the upper left panel), while the
absolute yield is at most $\SI{3}{\percent}$ of that of hadronic interactions.
Thus, we conclude that UPCs provide a negligible contribution to the amount of
$\pizero$s giving background photon events.
The $z$ distribution of $\pizero$s from UPCs (the upper right panel) is slightly
steeper than for hadronic interactions, but the absolute yield is once again
negligible.
On the other hand, neutrons produced in UPCs compete with those from hadronic
interactions at $\pt\lesssim\gev{0.2}$. This can be explained by the same
mechanism found in Fig.~\ref{fig:lhc}, namely dominance of the $\gamp \to
\pi^{+} + n$ channel via baryon resonances in UPCs. The $z$ distribution of
neutrons has a similar shape to that shown in the bottom right panel of
Fig.~\ref{fig:lhc}.
Concerning the measurements of forward prompt photons, $\pizero$s, and neutrons
in polarized $\pAu$ collisions, which are sensitive to the origin of spin
asymmetry~\cite{Kang}, we confirm that the UPC contribution to the total number
of prompt photons and $\pizero$s is negligible, while neutrons instead, are
produced in similar amounts from both UPCs and hadronic interactions.

We need to test now, both for the LHC and RHIC simulations, a possible
dependence of the $\pt$ and $z$ distributions on the cut off energy $\ecut$ that
was introduced in Sect.~\ref{sec:lowe} and \ref{sec:highe}.
The comparison of the two distributions at the LHC, with one based on the UPC
simulations with \textsc{dpmjet} above
$E_\gamma^{\textrm{cut},\textsc{dpmjet}}=\gev{6}$ and the other based on the UPC
simulations with \textsc{dpmjet} above
$E_\gamma^{\textrm{cut},\textsc{dpmjet}}=\gev{55}$, shows a negligible
difference between these two.
The comparison in the RHIC case shows larger differences than at LHC;
nevertheless they are at the same level of those between \textsc{dpmjet} and
\textsc{pythia}.
There is also a significant difference for the $\pt$ distributions of neutrons
at the LHC between \textsc{dpmjet} and \textsc{pythia} (the bottom left panel of
Fig.~\ref{fig:lhc}); \textsc{pythia} predicts a harder spectrum and has $\sim 8$
times larger number of neutrons at $\pt\approx\gev{2}$ than \textsc{dpmjet}.
This difference is mostly caused by a strong dependence of the multiplicity of
leading baryons on the minimum $\pt$ ($\ptmin$) for the multiple interactions
(i.e., hard scattering) implemented by each model. In general, a lower $\ptmin$
provides more multiple interactions in an event which then leads to a higher
multiplicity.
Changing the $\ptmin$ value in \textsc{pythia} (\texttt{PARP(81)}, \gev{1.9} as
a default) to the default value of \textsc{dpmjet} (\gev{2.5}) significantly
modifies the $\pt$ distribution in \textsc{pythia} bringing an overall agreement
between the two.

Looking back at Table~\ref{tbl:crosssection} we see that the effective cross
sections for $\pizero$ and neutron productions are defined as the cross sections
having at least one $\pizero$ or neutron hitting within the rapidity ranges
highlighted above. We find that UPC cross sections are similar or larger than
the hadronic cross sections, except for the $\pizero$ one at RHIC.
It should be noted that the effective cross section estimation involves
simulations of specific hadron productions and is thus no longer independent of
the generators used for hadronic interactions. The values in
Table~\ref{tbl:crosssection} for UPCs are calculated by using
\textsc{starlight}, \textsc{sophia}, and \textsc{dpmjet}, and those for hadronic
interactions are calculated by using \textsc{dpmjet}. Nevertheless, the essence
of our conclusions will not depend on the choice of event generators.

Finally, we roughly estimate the contribution to UPCs where the proton acts as
the photon source, which was not taken into account in the MC simulation of this
paper, and the possible changes to the $\pt$ and $z$ distributions. Because of
the directions of the colliding photon and nucleus and because of the smaller
cross sections (calculated in Sect.~\ref{sec:crosssection}), the number of
$\pizero$s and neutrons produced at $\ylab>3$ is negligible compared with that
from UPCs where the nucleus acts as the photon source.
Conversely, the two-photon exchange process followed by the $\gamp$ interaction,
currently not implemented in the MC simulation as well, would yield $\pizero$s
and neutrons at positive forward rapidities amounting to $\SI{15}{\percent}$
of those shown in Fig.~\ref{fig:lhc} at the LHC and to $\SI{9}{\percent}$ of
those shown in Fig.~\ref{fig:rhic} at RHIC.
A more detailed study would need to take two-photon and di-electron interactions
into account in the MC simulation framework and is thus beyond the scope of the
present paper.
 
%
%---- Reduction ----
%
\subsection{Reduction of ultra-peripheral collisions contributions}
\label{sec:reduction}

As seen in Fig.~\ref{fig:rapidity}, particle production in UPCs is clustered at
positive forward rapidity regions, whereas those in hadronic interactions show a
plateau at mid-rapidity and have also a large number of spectator nucleons at
$\etalab \sim -8$.
In this section, we present the two methods to separate UPCs and hadronic
interactions by exploiting those differences in the rapidity distributions.

First, we investigate the effects of requiring some activities in the
mid-rapidity region. The following cuts are applied to inclusive $\pizero$ and
neutron measurements to eliminate UPC induced events as well as to ensure that
the hadronic interaction events remain unchanged: (1) the number of charged
particles should be greater than 2, (2) the charged particles should have $\pt >
\gev{0.2}$, and (3) the charged particles should have rapidity $|\etalab|<2.5$
at the LHC and $|\etalab|<0.35$ at RHIC. The rapidity regions used in the cuts
correspond to the rapidity ranges of the ATLAS and PHENIX inner
detectors~\cite{ATLASInner,PHENIXInner}.

\begin{figure}[tbp]
  \centering
  \includegraphics[width=8.5cm, keepaspectratio]{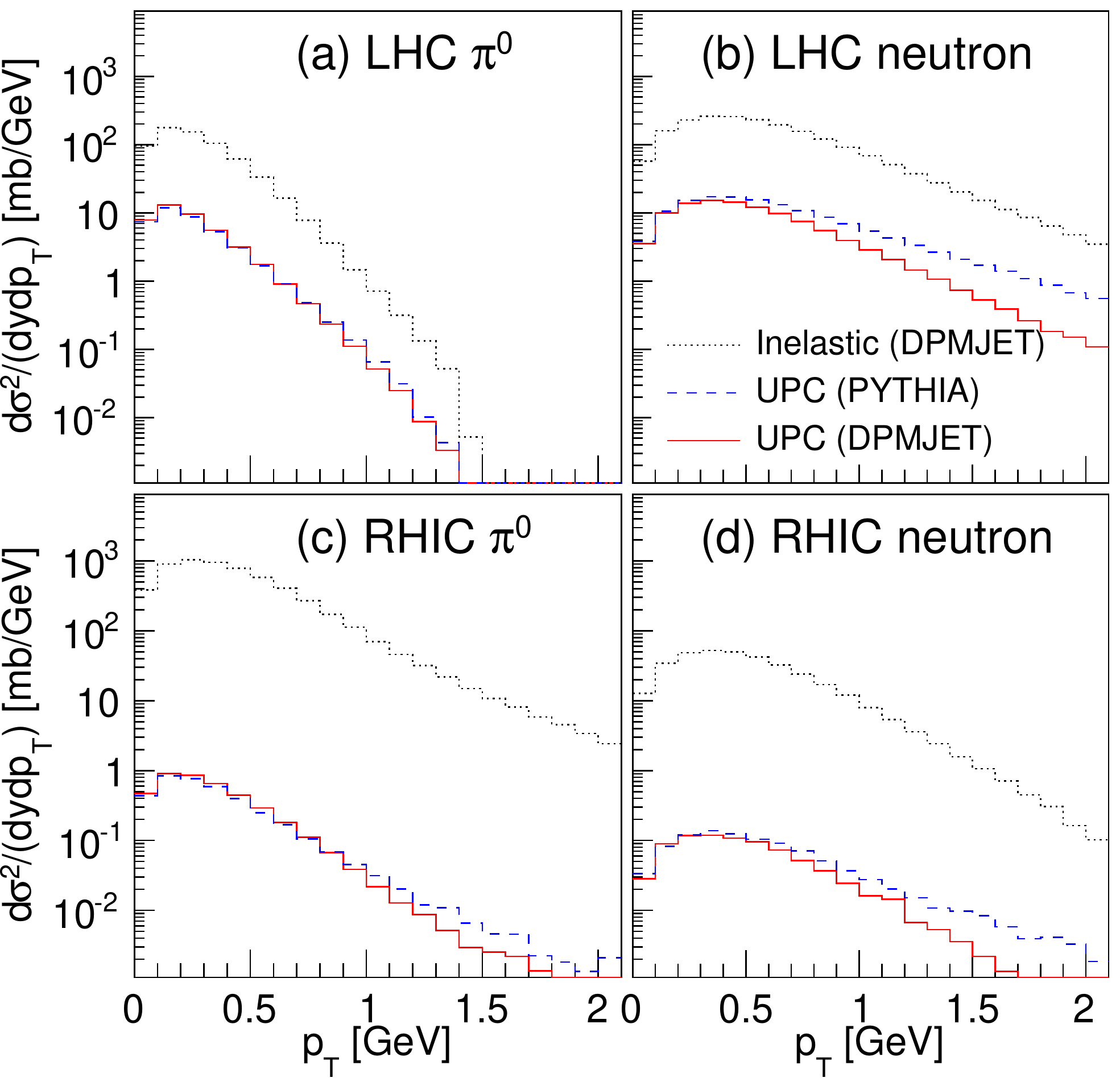}
  \caption{Simulated $\pt$ spectra for $\pizero$s and neutrons with the cuts
  applied at the LHC and RHIC (see text in detail). The solid curves and dashed
  curves indicate the UPC simulation events generated by using
  \textsc{starlight+sophia+dpmjet} and \textsc{starlight+sophia+pythia},
  respectively. The dotted curves indicate the simulated $\pA$ inelastic events
  with \textsc{dpmjet}.}
  \label{fig:bgcut1}
\end{figure}

Figure~\ref{fig:bgcut1} shows the $\pt$ distributions after the cuts at the LHC
and RHIC. The absolute yields of UPCs at the LHC (RHIC) are reduced to less than
$\SI{60}{\percent}$ ($\SI{8}{\percent}$) for $\pizero$s and $\SI{40}{\percent}$
($\SI{10}{\percent}$) for neutrons, even though the absolute yields of hadronic
interactions are kept to be larger than, at most, $\SI{85}{\percent}$
($\SI{50}{\percent}$) for $\pizero$'s and $\SI{85}{\percent}$
($\SI{20}{\percent}$) for neutrons.
Thus these cuts on the charged particles at mid-rapidity reduce the relative
yields of UPCs to hadronic interactions by, at most, $\SI{15}{\percent}$
($\SI{0.2}{\percent}$) for $\pizero$s and $\SI{25}{\percent}$
($\SI{1.5}{\percent}$) for neutrons.
However the cuts still leave a remnant contribution from UPCs and unavoidably
reduce hadronic interaction events. It should also be noted that the rejected
hadronic interaction events are mostly single and double diffractive events,
generally characterized by a small number of charged particles at mid-rapidity,
and that the reduction efficiency of $\SI{10}{\percent}-\SI{50}{\percent}$
highly depends on forward $\pizero$ and neutron energies.

Next, we test sharper cuts, which require an activity in the negative very
forward region, i.e., the direction of the nucleus remnant, and which can be
tagged by a ZDC. In this rapidity region, only hadronic interactions produce a
large number of spectator nucleons fragmented from the colliding nucleus. Each
nucleon has an approximate energy of \tev{1.58} at the LHC and \gev{100} at
RHIC.
The cuts applied to the simulated events consist of three requirements:
(1) the number of neutrons should be greater than 1 (note that a proton can be
swept away by the magnets located between an interaction point and a detection
point, and thus no proton reaches the detector), (2) the neutrons should have $E
> \tev{1}$ at the LHC and $E > \gev{50}$ at RHIC, and (3) the neutrons should
have a rapidity $\etalab<-6.5$. The rapidity region $\etalab<-6.5$ roughly
corresponds to the acceptance of the ZDC located in the nucleus-going side.
As was inferred in Fig.~\ref{fig:rapidity}, we find that the contribution of
UPCs is efficiently eliminated by these cuts, keeping the number of hadronic
interactions unchanged.

We conclude that a reduction in the contributions from UPCs to the measurements
of forward hadrons is certainly feasible by requiring some activity in mid and
forward rapidity regions. In particular, we expect a strong reduction if we
detect spectator nucleons, for example, with a ZDC, at negative rapidity
$\etalab \lesssim -6.5$.

%
%---- Conclusions ----
%
\section{Conclusions}
\label{sec:conclusions}

Hadron production in forward rapidity regions by ultra-peripheral $\pPb$
collisions at the LHC and $\pAu$ collisions at RHIC was discussed.
The present paper was based on MC simulations of the interaction of virtual
photons emitted by a fast moving nucleus with a proton, by using several event
generators: \textsc{starlight}, \textsc{sophia}, \textsc{dpmjet}, and
\textsc{pythia}. \textsc{starlight} in this paper was customized in order to
transfer the information on the simulated virtual photon to \textsc{sophia} and
to receive the information on the produced particles from \textsc{sophia}.
This modification made possible the simulation of UPC induced events starting
from the photopion production threshold.
We found large cross sections for $\pizero$ and neutron productions in the very
forward region, leading to a substantial background contribution to the
measurements of hadronic interactions at both the LHC and RHIC.
Therefore, the presence of UPCs had to be taken into account in the analyses
focused on the investigation of collective nuclear effects and spin physics, in
order to correctly evaluate the fraction of the hadronic cross section relative
to the measured cross section.
We propose two types of cuts to reduce the fraction of UPCs relative to hadronic
interactions; one requires the presence of charged particles at mid-rapidity,
while the other requires the presence of spectator neutrons at negative very
forward rapidity.
The former cut certainly reduces the UPC induced events while unavoidably
rejecting some hadronic events.
The second cut efficiently eliminates a fraction of forward $\pizero$s and
neutrons produced by UPCs and does not change the number of hadronic interaction
events.
The proposed methods in this paper were simply based on a tracking detector and
a ZDC and thus are generally applicable to the measurements in other experiments
that have a similar detector design.

%
%---- Acknowledgments ----
%
\begin{acknowledgements}
The author would like to thank the authors of \textsc{starlight},
\textsc{sophia}, \textsc{pythia}, \textsc{dpmjet}, and \textsc{crmc} for
providing the codes of their event generators.
The author is also grateful to \\R. D'Alessandro, K. Kasahara, Y. Muraki, and T.
Sako for the useful comments and discussions regarding this topic.
The author is supported by the Postdoctoral Fellowships for Research Abroad,
Japan Society for the Promotion of Science.
Part of this work was performed using the computer resources provided by CERN
and INFN-CNAF.
\end{acknowledgements}

%
%---- Bibliography ----
%

\end{document}